\documentstyle[prl,aps,twocolumn,amssymb,graphicx]{revtex}

\newlength{\charwidth}
\renewcommand\slash[1]{\settowidth{\charwidth}{$#1/$}#1\kern-.5\charwidth/}
\newcommand{\be}{\begin{equation}}
\newcommand{\ee}{\end{equation}}
\newcommand{\beqa}{\begin{eqnarray}}
\newcommand{\eeqa}{\end{eqnarray}}
\newcommand{\bean}{\begin{eqnarray*}}
\newcommand{\eean}{\end{eqnarray*}}

\renewcommand\d{{\mathrm d}}
\newcommand\D{{\cal D}}

\begin{document}
\title{Phase transition and critical behaviour of the d=3 Gross-Neveu model}
\author{F.~H\"of\/ling, C.~Nowak and C.~Wetterich}
\address{
Institut f\"ur Theoretische Physik, Universit\"at
Heidelberg, Philosophenweg 16, D-69120, Heidelberg Germany}

\maketitle

\small{}

\begin{abstract}
A second order phase transition for the three dimensional
Gross-Neveu model is established for one fermion species
\mbox{$N=1$}. This transition breaks a parity-like discrete
symmetry. It constitutes its peculiar universality class with critical exponent
$\nu = 0.63$ and scalar and fermionic anomalous dimension
$\eta_\sigma = 0.31$ and $\eta_\psi = 0.11$, respectively. We also
compute critical exponents for other $N$. Our results are based on
exact renormalization group equations.
\end{abstract}

\narrowtext


\bigskip
An understanding of systems with many fermionic degrees of freedom
is one of the big challenges in statistical physics. Due to the
anticommuting nature of the variables numerical simulations are
not straightforward---analytical methods are crucially needed. One
typically has to solve a functional integral for a $d$ dimensional
system with Grassmann variables. Approximate solutions for ``test
models'' would be of great value. The Gross-Neveu (GN) model
\cite{gn} is one of the simplest fermionic models. In three
dimensions a discrete symmetry forbids a mass term unless it is
spontaneously broken. In the symmetric phase the GN~model is
therefore a realization of a statistical system of gapless
fermions. For a large number $N$ of fermion species it is known
\cite{Gat,Hands,Reisz,Babaev} that a second order phase transition
separates the symmetric phase from an ordered phase where the
symmetry is spontaneously broken and the fermions become massive.
Using methods based on an exact renormalization group equation
\cite{vitale} a second order transition for $N \ge 2$ was
confirmed. We know, however, of no previous work which clarifies
the existence and nature of the phase transition in the simplest
model with only one fermion species. The model with one fermion
species is inaccessible to lattice simulations due to the fermion
doubling problem and the $1/N$ expansion is not expected to give
reasonable results for $N=1$. The case $N=1$ is also of special
interest since an order parameter $\langle
\bar\psi_j\,\psi^j\rangle \ne 0$ leads to a ground state which
does not admit any discrete symmetry involving the reflection of
all coordinates, in contrast to the models with $N \geq 2$.

In this letter we improve the exact renormalization group approach
and establish a second order phase transition for $N=1$. We also
compute the critical exponents. This is important beyond a
possible relevance for real physical systems: the GN~model
constitutes a peculiar universality class due to the presence of
massless fermions at the critical point. Just as the
$O(N)$-Heisenberg models for bosons, the GN~model could in the
future become a benchmark for our understanding of critical
systems in presence of fermions.

The GN~model describes $N$ fermionic fields with local quartic
interaction. Here $\psi_j,\, j=1...N$, are irreducible representations of the
group $O(d)$ including parity reflections, i.\,e. $2^{d/2}$ component Dirac spinors
for $d$ even and $2^{(d-1)/2}$ for $d$ odd. The classical Euclidean action
\begin{equation}
S=\int\!\d ^dx\,\Bigl\{\bar\psi_j(x)\,i\slash\partial\,\psi^j(x)
  + \frac{\bar G}{2}\bigl(\bar\psi_j(x)\,\psi^j(x)\bigr)^2 \Bigr\}.
\label{eq:GN-action}
\end{equation}
is symmetric under a coordinate reflection
$\psi(x) \mapsto -\psi(-x)\,,\,\bar \psi(x) \mapsto
\bar\psi(-x)$. (We note that $\psi$ and $\bar\psi$ are independent
variables in an Eucidean formulation.) A nonvanishing expectation value of $\bar \psi_j
\psi^j$ spontaneously breaks this symmetry. If the spinors
$\psi_j$ contain more than one irreducible representation of the
rotation (or Lorentz) group $SO(d)$ we can find alternative
definitions of the coordinate reflection where $\bar \psi_j
\psi^j$ remains invariant. In particular, this is realized for
Dirac spinors in even dimensions where a mass term couples two
irreducible representations $\psi_L$ and $\psi_R$. One may then
define a standard parity transformation under which $\bar \psi
\psi$ is invariant. In this case, however, $\bar \psi_R \psi_L \ne
0$ breaks a discrete chiral symmetry $\psi_L \mapsto
-\psi_L\,,\,\bar \psi_L \mapsto -\bar\psi_L$. We will concentrate
here on one two-component spinor in three dimensions. For $N=1$
the above ``parity transformation'' is the only possible choice of
coordinate reflections and $\bar\psi_j \psi^j \ne 0$ spontaneously
breaks this symmetry.

Depending on the value of the coupling $\bar G$ the model will be in
the symmetric phase or exhibit spontaneous symmetry breaking (SSB)
with a nontrivial expectation value of the order parameter
$\langle G\bar\psi_j\,\psi^j\rangle$. We will describe a space dependent fermion
bilinear $-i\,\bar G\bar\psi_j(x)\,\psi^j(x)$ by a real scalar field
$\sigma(x)$ such that SSB is indicated by nonzero homogeneous
$\langle\sigma\rangle \ne 0$. By a partial bosonization we express the GN~model
(\ref{eq:GN-action}) as an equivalent Yukawa model with
\begin{equation}
S_\sigma=\int\!\d ^dx\,\Bigl\{\bar\psi_j i\slash\partial\,\psi^j
  + i\sigma\bar\psi_j\psi^j + \frac{1}{2\bar G} \sigma^2 \Bigr\}.
\label{eq:sigma-action}
\end{equation}
The equivalence of the partition function can be seen by performing
the Gaussian $\sigma$-integration ($\bar\eta\psi$ means
$\int\!\d^dx\,\bar\eta_j(x)\,\psi^j(x)$ etc.)
\begin{eqnarray} Z[\bar\eta,\eta]&=&
  \int\!\D\sigma\D\psi\D\bar\psi\exp(-S_\sigma[\sigma,\psi,\bar\psi]
   + \bar\eta\psi + \eta\bar\psi) \nonumber\\
&=&\int\!\D\psi\D\bar\psi\exp\bigl(- \bar\psi i\slash\partial\,\psi
  +\bar\eta\psi + \eta\bar\psi - \frac{\bar G}{2}(\bar\psi\psi)^2\bigr) \\
&& \times \int\!\D\sigma\exp\bigl(-\frac{1}{2\bar G}
    (\sigma + i\bar G\bar\psi\psi)^2 \bigr),
  \nonumber
\end{eqnarray}
where the last factor yields an irrelevant constant.

A powerful tool for non-perturbative examinations are exact
renormalization group equations for the effective action $\Gamma$, i.\,e. the
generating functional of the 1PI Green's functions \cite{wett1,wett2}. Starting
with the classical action $S$ at the UV cutoff $\Lambda$, we obtain a type of
coarse-grained free energy $\Gamma_k$ by integrating out the quantum fluctuations
with momenta larger than a given scale $k$. Eventually, we reach the macroscopic
thermodynamic potential \cite{wett2} $\Gamma$ at $k\to 0$. The IR cutoff $k$
is implemented in a smooth way by
introducing a mass-like term $\Delta S_k$ into the classical action which
gives extra masses to modes with momenta smaller than $k$. In the
limit $k\to 0$ the IR cutoff is absent, and fluctuations at all scales have been
taken into account. At $k\to\Lambda$ all fluctuations are suppressed and
$\Gamma_\Lambda$ approaches $S$.

This procedure should be explained more precisely. For notation purposes we
combine the fermionic and (real) bosonic fields to a column vector
$\chi=(\sigma,\psi,\bar\psi^T)$, and the row $\cal J$ contains all the external sources
${\cal J}=(J,\bar\eta,\eta^T)$. We start with the generating functional for
the connected correlation functions in the presence of the IR cutoff:
\begin{equation} W_k[{\cal J}]=\ln\int\!\D\chi
 \exp\left(-S_\sigma[\chi] - \Delta S_k[\chi]+{\cal J}\chi\right)
\end{equation}
For vanishing $\Delta S_k$ (at $k\to 0$) this matches exactly with the free
energy of the Gross-Neveu model (\ref{eq:GN-action}). The effective action is
then defined via a modified Legendre transformation by
\begin{eqnarray}
\Gamma_k[\Phi]:=-W_k[{\cal J}[\Phi]] + {\cal J}[\Phi]\Phi - \Delta S_k[\Phi],
\end{eqnarray}
which depends on the expectation values of the fields $\Phi=\langle\chi\rangle$.

The infrared cutoff takes the form $\Delta S_k[\Phi]=\frac{1}{2}\Phi^T
R_k \Phi$, and $R_k$ is a matrix:
\begin{displaymath}
R_k(p,q)=\left(\begin{array}{ccc}
 R_{kB}(q)&0&0\\
 0&0&-R_{kF}^T(-q)\\
 0&R_{kF}(q)&0
\end{array}\right)(2\pi)^d \delta^d(p-q)
\end{displaymath}
with $R_{kB}(q)=Z_{\sigma,k}q^2 r_{kB}(q)$ and $R_{kF}(q)=-Z_{\psi,k}\slash q
r_{kF}(q)$. ($Z_{\sigma,k}$ and $Z_{\psi,k}$ are wavefunction renormalizations.)
With these definitions, an exact renormalization group equation for the scale
dependence of $\Gamma_k$ can be found \cite{wett1,wett2,Aoki}:
\begin{equation}
\partial_t \Gamma_k = \frac{1}{2}\mathrm{STr}
\left\{(\Gamma_k^{(2)}+R_k)^{-1}\,\partial_t R_k \right\}
\label{eq:erge}
\end{equation}
where $t=\ln(k/\Lambda)$. The super-trace runs over momenta and all internal
indices and provides appropriate minus signs for the fermionic sector. The heart
of the flow equation is the fluctuation matrix
\begin{equation}
\left(\Gamma_k^{(2)}(p,q)\right)_{ab} :=
\frac{\stackrel{\overrightarrow{}}{\delta}}{\delta\Phi_a^T(-p)}   \, \Gamma_k \,
  \frac{\stackrel{\overleftarrow{}}{\delta}}{\delta\Phi_b(q)}.
\label{eq:gamma2}
\end{equation}
Together with $R_k$ it represents the exact inverse propagator at the scale $k$.

Equation (\ref{eq:erge}) is an exact but complicated functional
differential equation. There is no way around some approximation by truncating
the most general form of $\Gamma_k$. We work here in the lowest order of a systematic
derivative expansion where $\Gamma_k$ contains a scalar potential, kinetic terms
and a Yukawa coupling. In momentum space it is given by ($\int\!\d q =
\int\!\d^d q/(2\pi)^d$):
\begin{eqnarray}
&&\Gamma_k[\sigma,\psi,\bar\psi] = \int\!\d^d x\,U_k(\sigma) + \int\!\d q\,
  \Bigl\{\frac{Z_{\sigma,k}}{2}\sigma(-q)\,q^2\sigma(q) \nonumber \\
&& - Z_{\psi,k} \bar\psi_j(q)\,\slash q\, \psi^j(q)
   + \int\!\d p\,i\bar h_k\bar\psi_j(p)\,\sigma(p-q)\,\psi^j(q) \Bigr\}
\label{eq:gammak}
\end{eqnarray}
The connection between $\bar G$ and $\bar h$ becomes clear if we rescale $\sigma$
in (\ref{eq:sigma-action}) to $\bar h\sigma$ and set $\bar G=\bar
h^2/m_\sigma^2$, $m_\sigma$ denoting the boson mass.

Using a truncation of the effective action causes the limit $k\to 0$ to depend
on the precise form of the cutoff functions. In order to take control of this
we have used two different choices, an exponential \cite{wett2} and a linear \cite{litim}
cutoff:
\begin{displaymath}
y r_B^{exp}(y)=\frac{y}{\exp(y)-1},\qquad y r_B^{lin}(y)=(1-y)\Theta(1-y),
\end{displaymath}
where $y=q^2/k^2$, and $r_F(y)$ is chosen in both cases such that $y(1+r_B)=y(1+r_F)^2$.
We introduce renormalized, dimensionsless quantities
$h^2_k=Z_\sigma^{-1}Z_\psi^{-2} k^{d-4} \bar{h}_k^2$,
${\tilde\rho}=\frac{1}{2}Z_\sigma k^{2-d}\sigma^2$,
$u_k=U_k k^{-d}$, and we use $u'_k=\frac{\partial u_k}{\partial \tilde\rho}$ etc.

We obtain a set of evolution equations for the effective parameters of the theory
by inserting (\ref{eq:gammak}) in the exact flow equation (\ref{eq:erge}). The evolution equation
of the effective scalar potential $u_k$ is found by evaluating $\Gamma_k^{(2)}$
for a constant scalar background field:
\beqa
\partial_t u_k &=& -d u_k
+ (d-2+\eta_\sigma){\tilde\rho}u'_k\nonumber\\
&+& 2 v_d \bigl\{l^d_0(u'_k+2{\tilde\rho}u''_k;\eta_\sigma)-
N' l^{(F)d}_0(2h_k^2{\tilde\rho};\eta_\psi)\bigr\},
\label{eq:flow_uk}
\eeqa
where we use $N'=d_\gamma N$, with $d_\gamma$ the dimension of the $\gamma$
matrices, and $v_d^{-1}=2^{d+1}\pi^{d/2}\Gamma(d/2)$.
The definition of the threshhold functions $l^d_n$ and $l^{(F)d}_n$
as well as $l^{(FB)d}_{n_1,n_2}$, $m^d_{n_1,n_2}$, $m^{(F)d}_{2/4}$, and
$m^{(FB)d}_{n_1,n_2}$ used below (see (\ref{eq:etaB})--(\ref{eq:flow_h2}))
is given in the appendix of \cite{wett2}. These functions contain all the momentum
integrations. For the linear cutoff the integrations can be performed
analytically---an enormous advantage for the subsequent numerics. We obtain:
\beqa
&&l^d_n(\omega;\eta_\sigma) = \frac{2(\delta_{n,0}+n)}{d}\Big(1
-\frac{\eta_\sigma}{d+2}\Big) \frac{1}{(1+\omega)^{n+1}}\nonumber\\
&&l^{(F)d}_n(\omega;\eta_\psi) = \frac{2(\delta_{n,0}+n)}{d}\Big(1
-\frac{\eta_\psi}{d+1}\Big) \frac{1}{(1+\omega)^{n+1}}\nonumber\\
&&l^{(FB)d}_{n_1,n_2}(\omega_1,\omega_2;\eta_\psi,\eta_\sigma) =
\frac{2}{d}\frac{1}{(1+\omega_1)^{n_1}(1+\omega_2)^{n_2}} \times\nonumber\\
&&\qquad\times \Big\{\frac{n_1}{1+\omega_1}
\left(1-\frac{\eta_\psi}{d+1}\right) + \frac{n_2}{1+\omega_2}
\left(1-\frac{\eta_\sigma}{d+2}\right)\Big\}\nonumber\\
&&m^d_{n_1,n_2}(\omega_1,\omega_2;\eta_\sigma)
  = \frac{1}{(1+\omega)^{n_1}(1+\omega)^{n_2}}\nonumber\\
&&m^{(F)d}_{2}(\omega;\eta_\psi) = \frac{1}{(1+\omega)^4}\quad,\quad
m^{(F)d}_{4}(\omega;\eta_\psi) = \frac{1}{(1+\omega)^4}+\nonumber\\
&&\qquad +\frac{1-\eta_\psi}{d-2}\frac{1}{(1+\omega)^3}
 - \left(\frac{1-\eta_\psi}{2d-4}+\frac{1}{4}\right)\frac{1}{(1+\omega)^2}\nonumber\\
&&m^{(FB)d}_{n_1,n_2}(\omega_1,\omega_2;\eta_\psi,\eta_\sigma)
 = \left(1-\frac{\eta_\sigma}{d+1}\right)\frac{1}{(1+\omega_1)^{n_1}(1+\omega_2)^{n_2}}\nonumber
\eeqa

The anomalous dimensions $\eta_\sigma$ and $\eta_\psi$ are defined as
\be
\eta_\sigma(k) = - \partial_t \ln Z_{\sigma,k}~,~~~
\eta_\psi(k) = - \partial_t \ln Z_{\psi,k}~.
\ee
Taking second derivatives of (\ref{eq:erge}) with respect to the fields
by means of (\ref{eq:gamma2}), we obtain evolution equations for the exact
inverse boson and fermion propagators. Expanding the momentum
dependence at $q^2=0$ for $\sigma=\sigma_{0k}$ at the potential minimum
yields equations for $\eta_\sigma$ and $\eta_\psi$:
\beqa
\eta_\sigma(k) &=& 8 \frac{v_d}{d} \Big\{ {\kappa}_k (3u''_k+2{\kappa}_k u'''_k)^2 m^d_{4,0}
(u'_k+2{\kappa}_k u''_k,0;\eta_\sigma)\nonumber\\
+ N' &h^2_k& \big[m^{(F)d}_{4}(2h^2_k {\kappa}_k;\eta_\psi)
- 2h^2_k {\kappa}_k m^{(F)d}_2(2h^2_k {\kappa}_k;\eta_\psi)\big]\Big\}
\label{eq:etaB}\\
\eta_\psi(k) &=& 8 \frac{v_d}{d} h^2_k m^{(FB)d}_{1,2}
(2h^2_k {\kappa}_k,u'_k+2{\kappa}_k u''_k;\eta_\psi,\eta_\sigma)
\label{eq:etaF}
\eeqa
Here ${\kappa}_k$ denotes the position of the minimum of $u_k(\tilde\rho)$ and
all derivatives of the potential are evaluated at $\tilde\rho=\kappa_k$.
Dividing the evolution equation of the fermion propagator by $\sigma$ and evaluating it
at zero momentum and $\sigma=\sigma_{0k}$ yields the evolution equation of $h^2_k$:
\beqa
\partial_t h^2_k &=& (2\eta_\psi + \eta_\sigma +d-4)h^2_k\nonumber\\
&+& 8 v_d h^4_k l^{(FB)d}_{1,1}(2h^2_k {\kappa}_k,u'_k+2{\kappa}_k u''_k;\eta_\psi,\eta_\sigma)
\label{eq:flow_h2}
\eeqa

We study the flow of the effective potential in two different ways that correspond to two
different truncations of $\Gamma_k$. First, we expand
the evolution equation of the potential (\ref{eq:flow_uk}) in a Taylor series up to
third order in $\tilde{\rho}$ around its minimum and evolve the resulting coupled ordinary
differential equations \cite{vitale}. In a second, more involved approach, the full equation
for $u'_k$ is discretized on a grid, yielding coupled ordinary differential
equations (one for each grid point) which are solved simultaneously. We use the equation for $u'_k$
instead of the one for $u_k$ for numerical reasons.

The only critical parameter in the theory is the four-fermion coupling $\bar G_{\Lambda cr}$.
This parameter has to be tuned in order to be near the phase transition. The flow
equations (\ref{eq:flow_uk})--(\ref{eq:flow_h2}) are evolved from an UV cutoff
scale $\Lambda$ to $k \rightarrow 0$. We concentrate on the three-dimensional case
with one fermion species ($d=3$ and $N=1$). The initial values of the parameters are chosen
such that $\Gamma_\Lambda = S_{\sigma}$: $Z_{\sigma\Lambda} = 10^{-10}\simeq 0$,
$Z_{\psi\Lambda}=1$, $\bar{h}^2_{\Lambda}=c\,\Lambda$ and
$u_{\Lambda}'(\tilde\rho)=e:=(Z_{\sigma\Lambda} \bar G_\Lambda \Lambda)^{-1}$
for all $\tilde\rho$. The GN~model corresponds to $c=1$ whereas for $c\ne 1$ we investigate
Yukawa type theories in the same universality class.

Very close to the phase transition we find scaling
solutions for all evolving parameters corresponding to vanishing beta functions.
This behaviour indicates a second order phase transition. The symmetric regime is
characterized by $u_k'(0)>0$ as well as $\kappa_k=0$,
while after spontaneous symmetry breaking $u_k'(0)$ becomes negative and
$\kappa_k>0$. Starting in the symmetric regime the system evolves into the SSB regime
and reaches the scaling solution. Further down the flow,
it evolves back into the symmetric regime for $\bar G_\Lambda < \bar G_{\Lambda_{cr}}$
or it remains in the SSB regime for $\bar G_\Lambda > \bar G_{\Lambda_{cr}}$. This behaviour
contrasts the one for $N \ge 2$ where the scaling solution is
located in the symmetric regime.
\begin{figure}
\centering
\includegraphics[angle=270,width=0.9\linewidth]{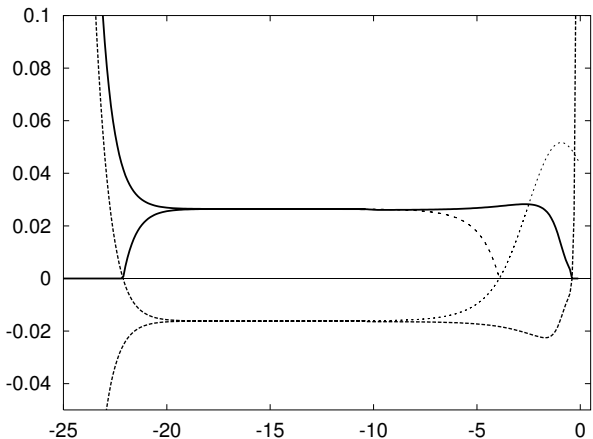}
\medskip
\caption{Critical flow of $10{\kappa}_k$ (full line) and $u'_k(0)$ (dashed line)
plotted as functions of $t=\ln(k/\Lambda)$. The thin dashed lines
show the flow for $c=10^{-10}$. We have chosen two initial values of $e$ very
near the phase transition. The flow separates only for very small $k$, according to the
respective phase.}
\label{fig:flow}
\end{figure}
Figure~\ref{fig:flow} shows the critical flow of $\kappa_k$ and $u'_k(0)$ for two different
values of $c$, namely $c = 1$ (full and dashed line, resp.) and $c = 10^{-10}$
(thin dashed lines). It nicely demonstrates that only the beginning of the flow
is affected by the choice of the non-critical parameters. All universal quantities
like critical exponents and mass gaps are independent of $c$. The polynomial expansion
of the potential works well only for small enough $c$. The value of the critical
coupling $\bar G_{\Lambda cr}$ depends, however, on the choice
of $c$. In the limit $c \rightarrow 1$ it converges towards a constant value,
which is reported in table~I.

We calculate the critical exponents characterizing a second order phase transition.
In order to test the reliability of the numerical algorithms, we determine the exponents
in the symmetric as well as in the ordered phase. The scale dependent renormalized boson mass $m^2_{\sigma R}(k)$ is defined as the curvature of the potential
at its minimum. In the SSB regime it is given by $m^2_{\sigma R}(k)=2 k^2{\kappa}_k u''_k({\kappa}_k)$,
in the symmetric regime by $m^2_{\sigma R}(k)=k^2 u'_k(0)$. In the ordered phase the running of
$m_{\sigma R}(k)$ essentially stops once $k$ becomes much smaller than
$m_{\sigma R}(k)$. The situation is slightly more difficult in the symmetric phase since the
fermions are massless. Their fluctuations induce a scale dependence of $Z_{\sigma,k}$ even for
very small momenta $k$: $\eta_\sigma \rightarrow 1$ for $k \rightarrow 0$ in contrast to
$\eta_\sigma \rightarrow 0$ for $k \rightarrow 0$ in the ordered phase.
To get rid of this problem we define the renormalized boson mass at some fixed small
ratio $r_c=k/\bar m_{\sigma R}$ \cite{vitale} as
\be
\bar m_{\sigma R}^2 = k_c^2\big(u'_{k_c}(0)-u'^{cr}_{k_c}(0)\big),\qquad k_c=r_c \bar m_{\sigma R}.
\ee
where $u'^{cr}_{k}(0)$ denotes $u'_k(0)$ on the critical trajectory.
In the numerical calculations we have used a ratio $r_c=0.01$, but our results do not
depend on $r_c$ for $r_c \lesssim 0.1$. Thus we define the critical exponent $\nu$
\beqa
 \nu = {1\over 2} \lim_{\delta e\rightarrow 0} {\partial \ln\bar m_{\sigma R}^2(\delta e) \over \partial\ln\delta e},
\label{nu}
\eeqa
where $\delta e=|e-e_{cr}|$. The exponent $\gamma$ is defined as usual, since the
unrenormalized boson mass $m^2_\sigma=Z_\sigma m_{\sigma R}^2$ is not affected by the
fluctuations of the massless fermions:
\beqa
\gamma = \lim_{\delta e\rightarrow 0}
 {\partial \ln m_\sigma^2(\delta e)\over \partial \ln\delta e}
\label{gamma}
\eeqa
In the ordered phase both exponents are defined as usual, using $m^2_{\sigma R}$ instead of
$\bar m^2_{\sigma R}$ for the definition of $\nu$. The critical exponent $\beta$ is defined as
\be
\beta={1\over 2}\lim_{\delta e\rightarrow 0} {\partial\ln\sigma_0^2 \over \partial\ln\delta e},
\ee
with $\sigma_0=\lim_{k\rightarrow 0} \sigma_{0k}$.
The exponents $\eta_\sigma$ and $\eta_\psi$ for the critical correlation functions are
computed by taking the values of the scale dependent anomalous dimensions $\eta_\sigma(k)$
and $\eta_\psi(k)$ at the scaling solution (sec.~4.2 of \cite{wett2}).
Table~I lists our results for the critical exponents. We find a good match of the values
in the two different phases. Besides, we have checked the index relations $\gamma=\nu(2-\eta_\sigma)$
and $\beta=\frac{1}{2}\nu(d-2+\eta_\sigma)$ which are well fulfilled. The dependence
of the exponents on the cutoff functions $r_B$ and $r_F$ as well as on the truncation can
also be seen from the table. The latter one seems to be weaker when using the linear cutoff.
The error on the exponent could be larger than the difference between the results for
different truncations of the potential and different cutoffs. This issue could be
investigated by extending the truncation (\ref{eq:gammak}), e.\,g. by including a
quartic fermion interaction $(\bar\psi\psi)^2$ along the lines discussed in \cite{gies}.

Since the fluctuations generate bosonic and fermionic masses, one might be interested
in the resulting gaps. They are proportional to the order parameter
$\rho_0=\frac{1}{2}Z_\sigma\sigma_0^2$:
\begin{equation}
m_{\sigma R} = \Delta_\sigma\rho_0, \qquad m_{\psi R}= \Delta_\psi\rho_0
\end{equation}
The gaps $\Delta_\sigma$ and $\Delta_\psi$ are shown in table~I.
\begin{center}
\medskip
\tabcolsep=2.5ex
\begin{tabular}{|c|c|c|c|c|}
\hline
Truncation & \multicolumn{2}{c|}{\hspace{2.5ex} full eq. for $u_k'$} &
 \multicolumn{2}{c|}{\hspace{1.5ex} Taylor expansion} \\
Cutoff & lin & exp & lin & exp \\
\hline \hline
$\nu_{symm}$         & 0.621 & 0.640 & 0.623 & 0.633 \\
$\gamma_{symm}$      & 1.051 & 1.077 & 1.053 & 1.062 \\
$\nu(2-\eta_\sigma)$ & 1.051 & 1.076 & 1.054 & 1.064 \\
$\nu_{ssb}$          & 0.620 & 0.637 & 0.622 & 0.632 \\
$\gamma_{ssb}$       & 1.050 & 1.071 & 1.053 & 1.062 \\
$\beta_{ssb}$        & 0.406 & 0.420 & 0.407 & 0.417 \\
$\frac{\nu}{2}
 (1+\eta_\sigma)$    & 0.405 & 0.420 & 0.407 & 0.417 \\
$\eta_\sigma$        & 0.308 & 0.319 & 0.308 & 0.319 \\
$\eta_\psi$          & 0.112 & 0.114 & 0.112 & 0.113 \\
\hline
$\Delta_\sigma$      & 16.0  & 17.6  & 16.8  & 18.1  \\
$\Delta_\psi$        & 14.2  & 14.9  & 14.4  & 15.2  \\
\hline
$\bar G_{\Lambda cr}
 \Lambda$            & 43.13 & 26.68 &       &       \\
\hline
\end{tabular}
\end{center}
\smallskip
\hspace{\parindent}TABLE I. Critical exponents and mass gaps for $N=1$ in three dimensions
\bigskip


We have also calculated the critical exponents in the
three-dimensional GN~model for $N>1$. They are compared with
results from other methods in table~II. For $N=2,4$ and $12$
critical exponents have been calculated in the $1/N$-expansion to
$O(1/N^2)$ \cite{gracey1,kaerk} and anomalous dimensions to
$O(1/N^3)$ using conformal techniques \cite{GraceyVasilev}.
Monte-Carlo methods have been used to calculate critical exponents
for $N=4$ which are compared with results from the $\epsilon =
4-d$ expansion to $O(\epsilon^2)$\cite{kaerk}. Our values for the
Taylor expansion with exponential cutoff (b) agree well with
\cite{vitale} where the precise numerical implementation was
different.

The largest discrepancy between different approaches concerns the
values of the anomalous dimensions. This is similar as for the
bosonic $O(N)$ models and is expected to improve substantially in
the next order in the derivative expansion \cite{seide}. The
overall picture is, however, quite satisfactory and lends support
to the validity of our method based on the exact renormalization
group equation. We believe that our finding of a second order
phase transition for $N=1$ is quite robust.

\begin{center}
\medskip
\tabcolsep=2.5ex
\begin{tabular}{|c|c|c|c|}
\hline
 N & 2 & 4 & 12 \\
\hline \hline
              & 0.927$^a$   & 1.018$^a$   & 1.018$^a$ \\
$\nu$         & 0.962$^b$   & 1.016$^b$   & 1.011$^b$ \\
              & 0.738$^c$   & 0.903$^c$   & 1.007$^c$ \\
              &             & 0.948$^d$   &           \\
              &             & 1.00(4)$^e$ &           \\
\hline
              & 0.525$^a$    & 0.756$^a$    & 0.927$^a$ \\
$\eta_\sigma$ & 0.554$^b$    & 0.786$^b$    & 0.935$^b$ \\
              & 0.635$^c$    & 0.776$^c$    & 0.914$^c$ \\
              &              & 0.763$^d$    & 0.913$^f$ \\
              &              & 0.754(8)$^e$ &           \\
\hline
              & 0.071$^a$ & 0.032$^a$ & 0.0087$^a$ \\
$\eta_\psi$   & 0.067$^b$ & 0.028$^b$ & 0.0057$^b$ \\
              & 0.105$^f$ & 0.044$^f$ & 0.0124$^f$ \\
\hline
$\Delta_\sigma$ & 15.1$^a$ & 12.0$^a$ & 5.5$^a$ \\
                & 17.3$^b$ & 14.0$^b$ & 6.4$^b$ \\
\hline
$\Delta_\psi$   & 10.8$^a$ & 7.5$^a$ &  3.3$^a$ \\
                & 11.4$^b$ & 7.9$^b$ &  3.4$^b$ \\
\hline
\end{tabular}
\end{center}
\smallskip
\hspace{\parindent}TABLE II. Critical exponents for various $N$ in three dimensions\\[1ex]
\mbox{\hspace{2em}} (a) linear cutoff, full equation for $u_k'$ \\
\mbox{\hspace{2em}} (b) exponential cutoff, Taylor expansion of $u_k'$ \\
\mbox{\hspace{2em}} (c) $1/N$-expansion (Pad\'e-Borel resummed) \\
\mbox{\hspace{2em}} (d) $\epsilon$-expansion (Pad\'e-Borel resummed) \\
\mbox{\hspace{2em}} (e) Monte-Carlo simulations\\
\mbox{\hspace{2em}} (f) conformal techniques
\bigskip

\end{document}